\pdfoutput=1
\documentclass[a4paper]{article}
\usepackage{graphicx}
\usepackage{mathtools}
\usepackage[margin=1in]{geometry}

\title{Non-parametric Bayesian modelling of digital gene expression data}
\author{Dimitrios V. Vavoulis\,$^{1,*}$ and Julian Gough\,$^{1,}$\footnote{to whom correspondence should be addressed} \\
\small{$^{1}$Department of Computer Science, University of Bristol, Bristol, United Kingdom} }
\date{January 7, 2013}

\begin{document}

\maketitle

\begin{abstract}
 
Next-generation sequencing technologies provide a 
revolutionary tool for generating gene expression data. Starting with a fixed RNA sample, 
they construct a library of millions of differentially abundant short sequence 
tags or ``reads'', which constitute a fundamentally discrete measure of
the level of gene expression. A common limitation in experiments using these technologies 
is the low number or even absence of biological replicates, which complicates the statistical 
analysis of digital gene expression data. Analysis of this type of data has often been based 
on modified tests originally devised for analysing microarrays; both these and even \textit{de novo} 
methods for the analysis of RNA-seq data are plagued by the common problem of low replication.

We propose a novel, non-parametric Bayesian approach for the analysis of digital gene 
expression data. We begin with a hierarchical model 
for modelling over-dispersed count data and a blocked Gibbs sampling algorithm 
for inferring the posterior distribution of model parameters conditional 
on these counts. The algorithm compensates for the problem of low numbers of biological replicates 
by clustering together genes with tag counts that are likely sampled from a common
distribution and using this augmented sample for estimating the parameters of this distribution.
The number of clusters is not decided \textit{a priori}, but it is inferred along with the remaining model 
parameters. We demonstrate the ability of this approach to model biological data with high fidelity 
by applying the algorithm on a public dataset obtained from cancerous and non-cancerous neural 
tissues.\footnote{Python code is available upon request. A user-friendly package is currently under development.}

\end{abstract}

\section{Introduction}

It is a common truth that our knowledge in Molecular Biology is only as good as
the tools we have at our disposal. Next-generation or high-throughput sequencing 
technologies provide a revolutionary tool in the aid of genomic studies by allowing 
the generation, in a relatively
short time, of millions of short sequence tags, which reflect particular aspects 
of the molecular state of a biological system. A common application of these technologies
is the study of the transcriptome, which involves a family of methodologies, including
RNA-seq (\cite{Wang:2009fk}), CAGE (Cap Analysis of Gene Expression; \cite{Shiraki:2003uq}) 
and SAGE (Serial Analysis of Gene Expression; \cite{Velculescu:1995kx}).
When compared to microarrays, this class of 
methodologies offers several advantages, including detection of a wider level of expression levels
and independence on prior knowledge of the biological system, which is required by hybridisation-based 
technologies, such as microarrays.  

Typically, an experiment in this category starts with the extraction of a snapshot RNA sample 
from the biological system of interest and its shearing in a large
number of fragments of varying lengths. The population of these fragments is then 
reversed-transcribed to a cDNA library and sequenced on a high-throughput 
platform, generating large numbers of short DNA sequences known as ``reads''. The 
ensuing analysis pipeline starts with mapping or aligning these reads on a reference genome. 
At the next stage, the mapped reads are summarised into gene-, exon- or transcript-level 
counts, normalised and further analysed for detecting differential gene expression (see
\cite{Oshlack:2010vn} for a review). 

It is important to realize that the normalised read (or tag) count data generated 
from this family of methodologies represents
the number of times a particular class of cDNA fragments has been sequenced, which is directly 
related to their abundance in the library and, in turn, the abundance of the associated
transcripts in the original sample. Thus, this count data is essentially 
a discrete or digital measure of gene expression, which is fundamentally different 
in nature (and, in general terms, superior in quality) from the continuous fluorescence 
intensity measurements obtained from the application of microarray technologies. Due to their 
better quality, next-generation sequence assays tend to replace microarray-based technologies,
despite their higher cost (\cite{Carninci:2009fk}).  

One approach for the analysis of count 
data of gene expression is to transform the counts to approximate normality and then 
apply existing methods aimed at the analysis of microarrays (see for example, 
\cite{t-Hoen:2008ys,Cloonan:2008zr}). However, as noted in \cite{McCarthy:2012ve},
this approach may fail in the case of very small counts (which are far 
from normally distributed) and also due to the strong mean-variance relationship of count data, 
which is not taken into account by tests based on a normality assumption. Proper statistical modelling and 
analysis of count data of gene expression requires 
novel approaches, rather than adaptation of existing methodologies, which aimed from
the beginning at processing continuous input.    
 
Formally, the generation of count data using next-generation sequencing assays can 
be thought of as random sampling of an underlying population of cDNA fragments. 
Thus, the counts for each tag describing a class of cDNA fragments can, in principle, 
be modelled using the Poisson distribution, whose variance is, by definition, equal 
to its mean. However, it has been shown that, in real count data of gene expression,
the variance can be larger that what is predicted by the Poisson distribution 
(\cite{Lu:2005dq,Robinson:2007bh,Robinson:2008cr,Nagalakshmi:2008qf}). An
approach that accounts for the so-called ``over-dispersion'' in the data
is to adopt quasi-likelihood methods, which augment the variance of the Poisson
distribution with a scaling factor, thus by-passing the assumption of equality between
the mean and variance (\cite{AueDoe11,Srivastava:2010nx,Wang:2010oq,Langmead:2010ly}). 
An alternative approach is to use the Negative Binomial distribution, which 
is derived from the Poisson, assuming a Gamma-distributed rate parameter. The Negative Binomial
distribution incorporates both a mean and a variance parameter, thus modelling over-dispersion
in a natural way (\cite{Anders:2010kl,Hardcastle:2010tg,McCarthy:2012ve}). An overview 
of existing methods for the analysis of gene expression count data can be found in \cite{Oshlack:2010vn}
and \cite{Kvam:2012hc}

Despite the decreasing cost of next-generation sequencing assays (and also
due to technical and ethical restrictions), digital datasets of gene expression
are often characterised by a small number of biological replicates or no replicates
at all. Although this complicates any effort to statistically analyse the data, 
it has led to inventive attempts at estimating as accurately as possible the biological
variability in the data given very small samples. One approach is to assume 
a locally linear relationship between the variance and the mean in the 
Negative Binomial distribution, which allows estimating the variance by pooling 
together data from genes with similar expression levels (\cite{Anders:2010kl}). 
Alternatively, one can make the rather restrictive assumption that all genes share the same 
variance, in which case the over-dispersion parameter in the Negative Binomial distribution 
can be estimated from a very large set of datapoints (\cite{Robinson:2007bh}). 
A further elaboration of this approach is to 
assume a unique variance per gene and adopt a weighted-likelihood methodology 
for sharing information between genes, which allows for an improved estimation of 
the gene-specific over-dispersion parameters (\cite{McCarthy:2012ve}). Another yet distinct 
empirical Bayes approach is implemented in the software \textit{baySeq}, which adopts a form
of information sharing between genes by assuming the same prior distribution among
the parameters of samples demonstrating a large degree of similarity (\cite{Hardcastle:2010tg}).  
           
In summary, proper statistical modelling and analysis of digital gene expression 
data requires the development of novel methods, which take into account both the 
discrete nature of this data and the typically small number (or even the absence)
of biological replicates. The development of such methods is particularly urgent 
due to the huge amount of data being generated by high-throughput sequencing assays.
In this paper, we present a method for modelling digital gene expression data that utilizes
a novel form of information sharing between genes (based on non-parametric Bayesian clustering)
to compensate for the all-too-common problem of low or no replication, which plagues most 
current analysis methods.

\section{Approach}

We propose a novel, non-parametric Bayesian approach for the analysis of digital gene 
expression data. Our point of departure is a hierarchical model  
for over-dispersed counts. The model is built around the Negative Binomial distribution,
which depends, in our formulation, on two parameters: the mean and an over-dispersion
parameter. We assume that these parameters are sampled from a Dirichlet
process with a joint Inverse Gamma - Normal base distribution, which we have 
implemented using stick breaking priors. By construction, the model imposes a clustering
effect on the data, where all genes in the same cluster are statistically described 
by a unique Negative Binomial distribution. This can be thought of as a form of 
information sharing between genes, which permits pooling together data from genes 
in the same cluster for improved estimation of the mean and over-dispersion parameters,
thus bypassing the problem of little or no replication.
We develop a blocked Gibbs sampling algorithm for estimating the posterior distributions
of the various free parameters in the model. These include the mean and over-dispersion
for each gene and the number of clusters (and
their occupancies), which does not need to be fixed \textit{a priori}, as 
in alternative (parametric) clustering methods. In principle, the proposed method can be 
applied on various forms of digital gene expression data (including RNA-seq, CAGE, SAGE, Tag-seq, etc.) 
with little or no replication and it is actually applied on one such example dataset herein. 

\section{Modelling over-dispersed count data}

The digital gene expression data we are considering is arranged in an $M \times N$ 
matrix, where each of the $N$ rows corresponds to a different gene and each of 
the $M$ columns corresponds to a different sample. Furthermore, all samples are 
grouped in $L$ different classes (i.e. tissues or experimental conditions). 
It holds that $L \le M$, where the equality is true if there are no replicas in the 
data.

\begin{figure}[!tpb]
\centerline{\includegraphics[width=3.27in]{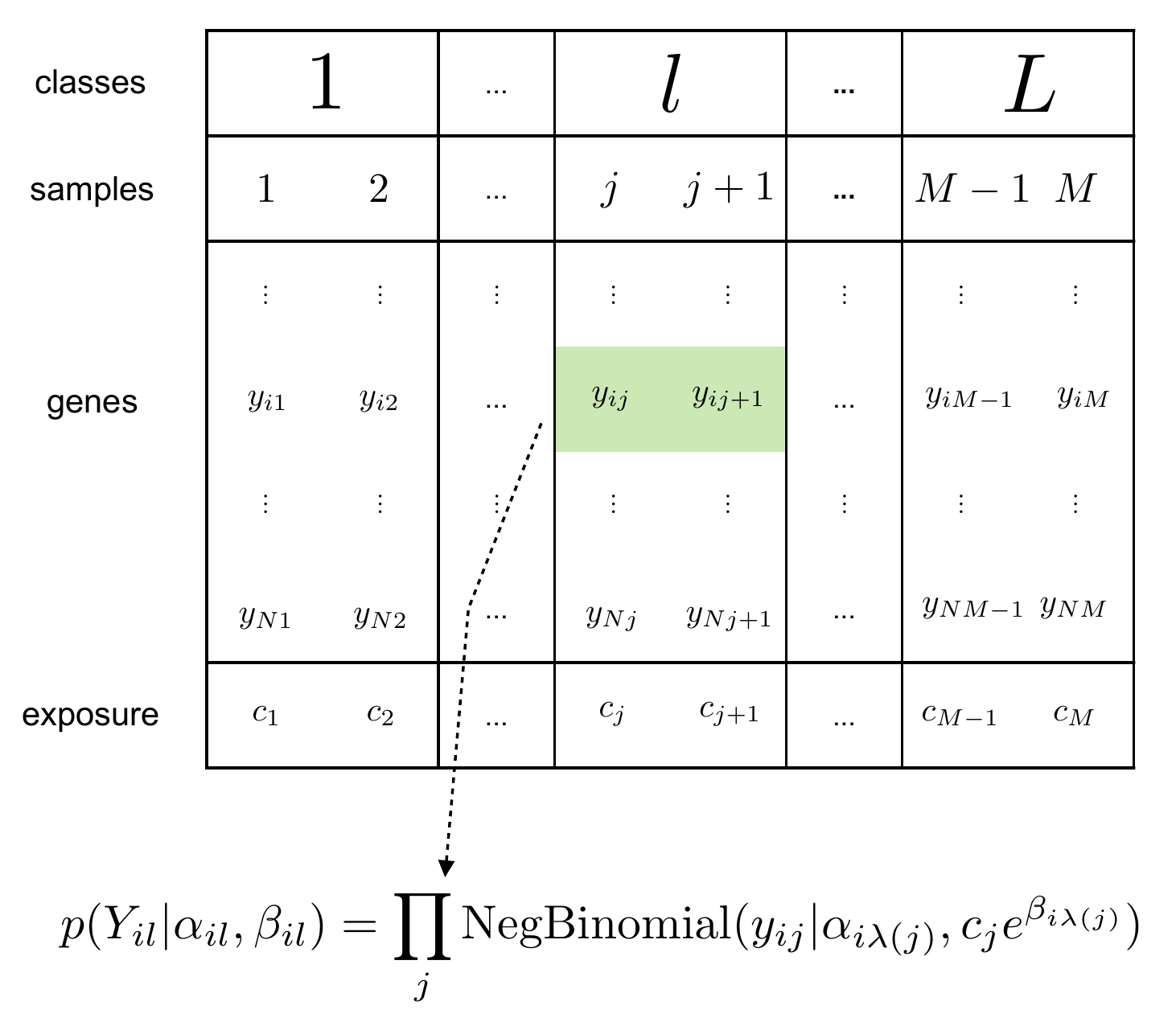}}
\caption{Format of digital gene expression data. Rows correspond to genes and
columns correspond to samples. Samples are grouped into classes (e.g.
tissues or experimental conditions). Each element of the data matrix
is a whole number indicating the number of counts or reads corresponding
to the $i^{th}$ gene at the $j^{th}$ sample. The sum of the reads
across all genes in a sample is the depth or exposure of that sample.\label{fig:data_matrix}}
\end{figure}

We indicate the number of reads for the $i^{th}$ gene at the $j^{th}$
sample with the variable $y_{ij}$. We assume that $y_{ij}$ is Poisson-distributed
with a gene- and sample-specific rate parameter $r_{ij}$.
The rate parameter $r_{ij}$ is assumed random itself and it is modelled
using a Gamma distribution with shape parameter $\alpha_{i\lambda(j)}$
and scale parameter $s_{ij}$.
The function $\lambda(\cdot)$ in the subscript of the shape parameter
maps the sample index $j$ to an integer indicating the class this
sample belongs to. Thus, for a particular gene and class, the shape
of the Gamma distribution is the same for all samples. Under this
setup, the rate $r_{ij}$ can be integrated (or marginalised) out,
which gives rise to the Negative Binomial distribution with parameters
$\alpha_{i\lambda(j)}$ and $\mu_{ij}=\alpha_{i\lambda(j)}s_{ij}$
for the number of reads $y_{ij}$:
\begin{equation}
y_{ij}|\alpha_{i\lambda(j)},\mu_{ij} \sim
\frac{\Gamma(y_{ij}+\alpha_{i\lambda(j)})}{\Gamma(\alpha_{i\lambda(j)})\Gamma(y_{ij}+1)}\left(\frac{\alpha_{i\lambda(j)}}{\alpha_{i\lambda(j)}+\mu_{ij}}\right)^{\alpha_{i\lambda(j)}}\left(\frac{\mu_{i\lambda(j)}}{\alpha_{i\lambda(j)}+\mu_{ij}}\right)^{y_{ij}}\label{eq:negbin}
\end{equation}
\noindent where $\mu_{ij}$ is the mean of the Negative Binomial distribution
and $\mu_{ij}+\alpha_{i\lambda(j)}^{-1}\mu_{ij}^{2}$ is the variance.
Since the variance is always larger than the mean by the quantity
$\alpha_{i\lambda(j)}^{-1}\mu_{ij}^{2}$, the Negative Binomial distribution
can be thought of as a generalisation of the Poisson distribution,
which accounts for over-dispersion. Furthermore, we model the mean
as $\mu_{ij}=c_{j}e^{\beta_{i\lambda(j)}}$, where the offset $c_{j}=\sum_{i=1}^{N}y_{ij}$
is the depth or exposure of sample $j$ and $\beta_{i\lambda(j)}$
is, similarly to $\alpha_{i\lambda(j)}$, a gene- and class-specific
parameter. This formulation ensures that $\mu_{ij}$ is always positive,
as it oughts to. 

Given the model above, the likelihood of observed reads $Y_{il}=\{y_{ij}:\lambda(j)=l\}$
for the $i^{th}$ gene in class $l$ is written as follows:
\begin{eqnarray}
p(Y_{il}|\alpha_{il},\beta_{il}) & = & \prod_{j}p(y_{ij}|\alpha_{i\lambda(j)},\beta_{i\lambda(j)})\nonumber \\
 & = & \prod_{j}\text{NegBinomial}(y_{ij}|\alpha_{i\lambda(j)},c_{j}e^{\beta_{i\lambda(j)}})\label{eq:negbinlik1}
\end{eqnarray}
\noindent where the index $j$ satisfies the condition $\lambda(j)=l$. By extension,
for the $i^{th}$ gene across all sample classes, the likelihood of
observed counts $Y_{i}=\{y_{ij}:\lambda(j)=l,l=1,\ldots,L\}$ is written
as: 
\begin{eqnarray}
p(Y_{i}|\alpha_{i1},\beta_{i1},\ldots,\alpha_{iL},\beta_{iL}) & = & \prod_{l}p(Y_{il}|\alpha_{il},\beta_{il})\label{eq:negbinlik2}
\end{eqnarray}
\noindent where the class indicator $l$ runs across all $L$ classes.

\subsection{Information sharing between genes}

A common feature of digital gene expression data is the small number
of biological replicates per class, which makes any attempt to estimate
the gene- and class-specific parameters $\theta_{il}=\{\alpha_{il},\beta_{il}\}$
through standard likelihood methods a futile exercise. In order to
make robust estimation of these parameters feasible, some form of
information sharing between different genes is necessary. In the present
context, information sharing between genes means that not all values
of $\theta_{il}$ are distinct; different genes (or the same gene
across different sample classes) may share the same values for these
parameters. This idea can be expressed formally by assuming that $\theta_{il}$
is random with an infinite mixture of discrete random measures as
its prior distribution:
\begin{equation}
\theta_{il}\sim\sum_{k=1}^{\infty}w_{k}\delta_{\theta_{k}^{*}},\quad0\le w_{k}\le1,\quad\sum_{k=1}^{\infty}w_{k}=1\label{eq:infmix}
\end{equation}
\noindent where $\delta_{\theta_{k}^{*}}$ indicates a discrete random measure
centered at $\theta_{k}^{*}=\{\alpha_{k}^{*},\beta_{k}^{*}\}$ and
$w_{k}$ is the corresponding weight. Conceptually, the fact that
the above summation goes to infinity expresses our lack of prior knowledge
regarding the number of components that appear in the mixture, other
than the obvious restriction that their maximum number cannot be larger
than the number of genes times the number of sample classes. 

In this formulation, the parameters $\theta_{k}^{*}$ are sampled
from a prior base distribution $G_{0}$ with hyper-parameters $\phi$,
i.e. $\theta_{k}^{*}|\phi\sim G_{0}(\phi)$. We assume that $\alpha_{k}^{*}$
is distributed according to an inverse Gamma distribution with shape
$a_{\alpha}$ and scale $s_{\alpha}$, while $\beta_{k}^{*}$ follows
the Normal distribution with mean $\mu_{\beta}$ and variance $\sigma_{\beta}^{2}$.
Thus, $G_{0}$ is a joint distribution as follows:
\begin{eqnarray}
\overbrace{\alpha_{k}^{*},\beta_{k}^{*}}^{\theta_{k}^{*}}|\overbrace{a_{\alpha},s_{\alpha},\mu_{\beta},\sigma_{\beta}^{2}}^{\phi} \sim 
\overbrace{\text{InvGamma}(a_{\alpha},s_{\alpha})\cdot\text{Normal}(\mu_{\beta},\sigma_{\beta}^{2})}^{G_{0}(\phi)},\qquad k=1,2,\ldots\label{eq:thetaprior}
\end{eqnarray}
\noindent Given the above, $\alpha_{k}^{*}$ can take only positive values,
as it oughts to, while $\beta_{k}^{*}$ can take both positive and
negative values. 

What makes the mixture in Eq.~\ref{eq:infmix} special is the procedure
for generating the infinite sequence of mixing weights. We set $w_{1}=V_{1}$
and $w_{k}=V_{k}\prod_{m=1}^{k-1}\left(1-V_{m}\right)$ for $k\ge2$,
where $\{V_{1},\ldots,V_{k}\}$ are random variables following the
Beta distribution, i.e. $V_{k}\sim\text{Beta}(a_{k},b_{k})$. This
constructive way of sampling new mixing weights resembles a stick-breaking
process; generating the first weight $w_{1}$ corresponds to breaking
a stick of length $1$ at position $V_{1}$; generating the second
weight $w_{2}$ corresponds to breaking the remaining piece at position
$V_{2}$ and so on. Thus, we write:
\begin{equation}
w_{k}|a_{k},b_{k}\sim\text{Stick}(a_{k},b_{k}),\qquad k=1,2,\ldots
\end{equation}
\noindent There are various ways for defining the parameters $a_{k}$ and \textbf{$b_{k}$}.
Here, we consider only the case where $a_{k}=1$ and $b_{k}=\eta$,
with $\eta>0$. This parametrisation is equivalent to setting the
prior of $\theta_{il}$ to a Dirichlet Process with base distribution
$G_{0}$ and concentration parameter $\eta$. By construction, this
procedure leads to a rapidly decreasing sequence of sampled weights,
at a rate which depends on $\eta$. For values of $\eta$ much smaller
than $1$, the weights $w_{k}$ decrease rapidly with increasing $k$,
only one or few weights have significant mass and the parameters $\theta_{il}$
share a single or a small number of different values $\theta_{k}^{*}$.
For values of the concentration parameter much larger than $1$, the
weights $w_{k}$ decrease slowly with increasing $k$, many weights
have significant mass and the values of $\theta_{il}$ tend to be
all distinct to each other and distributed according to $G_{0}$.
Below, we set $\eta=1$, which results in a balanced decrease of the
weight mass with increasing $k$. In particular, for $\eta=1$, $\log(w_{k})$
decreases (on average) in an unbiased manner with increasing $k$. 

\begin{figure}[!tpb]
\centerline{\includegraphics[width=3.27in]{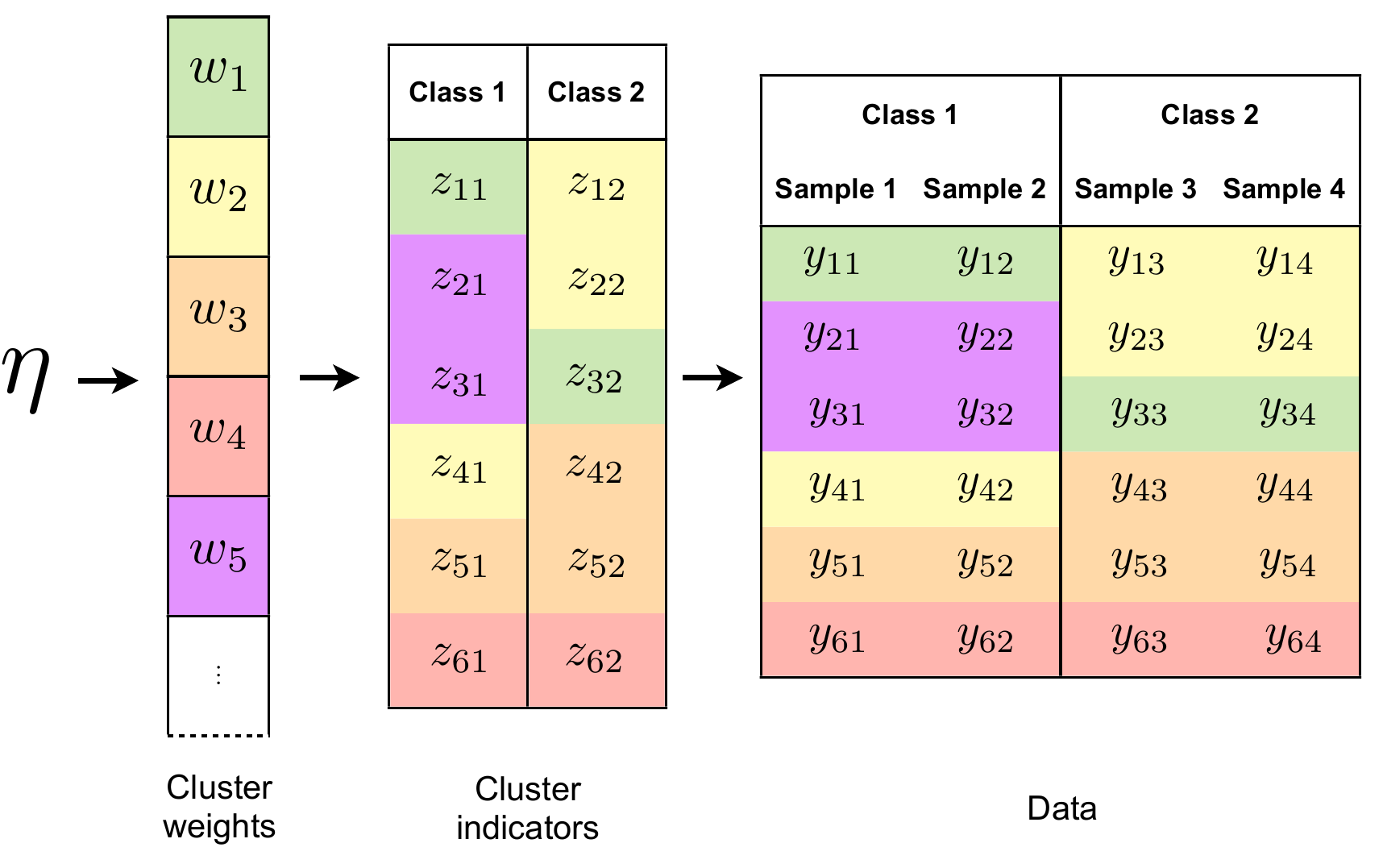}}
\caption{The clustering effect that results from imposing a stick-breaking
prior on the gene- and class-specific model parameters, $\theta_{il}$.
A matrix of indicator variables is used to cluster the observed count
data into a finite number of groups, where the genes in each group
share the same model parameters. The number of clusters is not known
\textit{a priori}. The distribution of weight mass among the various
clusters in the model is determined by parameter $\eta$.\label{fig:clustering}}
\end{figure}

Given the above formulation, sampling $\theta_{il}$ from its prior
distribution is straightforward. First, we introduce an indicator
variable $z_{il}\in\{1,2,\ldots\}$, which points to the value of
$\theta_{k}^{*}$ corresponding to the $i^{th}$ gene in class $l$.
We sample such indicator variables for each gene in each class from
the Categorical distribution, i.e. $z_{il}\sim\text{Categorical}(w_{1},w_{2,}\ldots)$,
and set $\theta_{il}\equiv\theta_{z_{il}}^{*}$. Although $G_{0}$
is continuous, the distribution of $\theta_{il}$ is almost surely
discrete and, therefore, its values are not all distinct. Different
genes may share the same value of $\theta^{*}$ and, thus, all genes
are grouped in a finite (unknown) number of clusters, according to
the value of $\theta_{k}^{*}$ they share. Modelling digital gene
expression data using this approach is one way to bypass the problem
of few (or the absence of) technical replicates, since the data from
all genes in the same cluster are pooled together for estimating the
parameters that characterise this cluster. The clustering effect described
in this section is illustrated in Fig. \ref{fig:clustering}.

\subsection{Generative model}

The description in the previous paragraphs suggests a hierarchical
model, which presumably underlies the stochastic generation of the
data matrix in Fig. \ref{fig:data_matrix}. This model is explicitly
described below: 
\begin{eqnarray}
\theta^*_k|a_\alpha,s_\alpha,\mu_\beta,\sigma^2_\beta &\sim& \text{InvGamma}(a_\alpha,s_\alpha)\cdot\text{Normal}(\mu_\beta,\sigma^2_\beta) \nonumber \\
w_1,w_2,\ldots|\eta &\sim& \text{Stick}(1,\eta) \nonumber \\
z_{i\lambda(j)}|w_1,w_2,\ldots &\sim& \text{Categorical}(w_1,w_2,\ldots)  \nonumber \\
\theta_{i\lambda(j)} &\equiv& \theta^*_{z_{i\lambda(j)}}     \nonumber \\
y_{ij}|\theta_{i\lambda(j)} &\sim& \text{NegBinomial}\left(\theta_{i\lambda(j)}\right) 
\end{eqnarray}

\noindent At the bottom of
the hierarchy, we identify the measured reads $y_{ij}$ for each gene
in each sample, which follow a Negative Binomial distribution with
parameters $\theta_{i\lambda(j)}=\{\alpha_{i\lambda(j)},\beta_{i\lambda(j)}\}$.
The parameters of the Negative Binomial distribution $\theta_{i\lambda(j)}$
are gene- and class-specific and they are completely determined by
an also gene- and class-specific indicator variable $z_{i\lambda(j)}$
and the centers $\theta_{k}^{*}$ of the infinite mixture of point
measures in Eq. \ref{eq:infmix}. These centers are distributed according
to a joint inverse Gamma and Normal distribution with hyper-parameters
$\phi=\{a_{\alpha},s_{\alpha},\mu_{\beta},\sigma_{\beta}^{2}\}$,
while the indicator variables are sampled from a Categorical distribution
with weights $\{w_{1},w_{2},\ldots\}$. These are, in turn, sampled
from a stick-breaking process with concentration parameter $\eta$.
In this model, $\phi$, $w_{k}$, $\theta_{k}^{*}$ and $z_{i\lambda(j)}$
are latent variables, which are subject to estimation based on the
observed data. 

\section{Inference}

At this point, we introduce some further notation. We indicate the
$N\times L$ matrix of indicator variables with the letter $Z$; $\Theta^{*}=\{\theta_{1}^{*},\theta_{2}^{*},\ldots\}$
lists the centers of the point measures in Eq. \ref{eq:infmix} and
$W=\{w_{1},w_{2},\ldots\}$ is the vector of mixing weights. 

We are interested in computing the joint posterior density $p(Z,W,\Theta^{*},\phi|Y)$,
where $Y$ is a matrix of count data as in Fig. \ref{fig:data_matrix}.
We approximate the above distribution through numerical (Monte Carlo)
methods, i.e. by sampling a large number of $\{\Theta^{*},W,Z,\phi\}$-tuples
from it. One way to achieve this is by constructing a Markov chain,
which admits $p(Z,W,\Theta^{*},\phi|Y)$ as its stationary distribution.
Such a Markov chain can be constructed by using Gibbs sampling, which
consists of alternating repeated sampling from the full conditional
posteriors $p(\Theta^{*}|Y,Z,\phi)$, $p(W|Z)$, $p(Z|Y,\Theta^{*},W)$
and $p(\phi|\Theta^{*},Z)$. Below, we explain how to sample from
each of these conditional distributions.

\subsubsection*{Sampling from the conditional posterior $p(\Theta^{*}|Y,Z,\phi)$}

In order to sample from the above distribution it is convenient to
truncate the infinite mixture in Eq. \ref{eq:infmix} by rejecting
all terms with index larger than $K$ and setting $w_{K}=1-\sum_{k=1}^{K-1}w_{k}$,
which is equivalent to setting $V_{K}=1$. It has been shown that
the error associated with this approximation when $V_{k}\sim\text{Beta}(1,\eta)$
is less than or equal to $4NM\exp(-\frac{K-1}{\eta})$ (\cite{Ishwaran2001}). For example,
for $N=14\times10^{3}$, $M=6$, $K=200$ and $\eta=1$, the error
is minimal (less than $10^{-80}$). Thus, the truncation should be
virtually indistinguishable from the full (infinite) mixture.

Next, we distinguish between $K_{ac}$ active clusters ($\Theta_{ac}^{*}$)
and $K_{in}$ inactive clusters ($\Theta_{in}^{*}$), such that $\Theta^{*}=\{\Theta_{ac}^{*},\Theta_{in}^{*}\}$
and $K=K_{ac}+K_{in}$. Active clusters are those containing at least
one gene, while those containing no genes are considered inactive.
We write:

\begin{eqnarray*}
p(\Theta^{*}|Y,Z,\phi) & = & p(\Theta_{ac}^{*},\Theta_{in}^{*}|Y,Z,\phi)\\
 & = & p(\Theta_{ac}^{*}|Y,Z,\phi)p(\Theta_{in}^{*}|\phi)
\end{eqnarray*}
Updating the inactive clusters is a simple matter of sampling $K_{in}$
times from the joint distribution in Eq. \ref{eq:thetaprior} given
the hyper-parameters $\phi$. Sampling the active clusters is more
complicated and involves sampling each active cluster center $\theta_{ac,k}^{*}$
individually from its respective posterior, $p(\theta_{ac,k}^{*}|Y_{ac,k})$,
where $Y_{ac,k}$ is a matrix of measured count data for all genes
in the $k^{th}$ active cluster. Sampling $\theta_{ac,k}^{*}=\{\alpha_{ac,k}^{*},\beta_{ac,k}^{*}\}$
is done using the Metropolis algorithm with acceptance probability:

\begin{equation}
P_{acc}=min\left(1,\frac{p(Y_{ac,k}|\theta_{ac,k}^{+})}{p(Y_{ac,k}|\theta_{ac,k}^{*})}\frac{p(\theta_{ac,k}^{+}|\phi)}{p(\theta_{ac,k}^{*}|\phi)}\right)\label{eq:metropolistheta}
\end{equation}
where the superscript $^{+}$ indicates a candidate vector of parameters.
Each of the two elements ($\alpha$ and $\beta$) of this vector is
drawn from a symmetric proposal of the following form:

\begin{equation}
q(x^{+}|x^{*})=x^{*}\exp(0.01\cdot r)\label{eq:propdist}
\end{equation}
where the random number $r$ is sampled from the standard Normal distribution,
i.e. $r\sim\text{Normal}(0,1)$. The prior of $p(\theta_{ac,k}^{*}|\phi)$
is a joint Inverse Gamma - Normal distribution, as shown in Eq. \ref{eq:thetaprior},
while the likelihood function $p(Y_{ac,k}|\theta_{ac,k}^{*})$ is
a product of Negative Binomial probability distributions, similar
to those in Eqs. \ref{eq:negbinlik1} and \ref{eq:negbinlik2}. 

\subsubsection*{Sampling from the conditional posterior $p(Z|Y,\Theta^{*},W)$}

Each element $z_{il}$ of the matrix of indicator variables $Z$ is
sampled from a Categorical distribution with weights $\pi_{il}=\{\pi_{il}^{1},\ldots,\pi_{il}^{K}\}$,
where $\pi_{il}^{k}=\Pi_{il}^{k}/\sum_{m=1}^{K}\Pi_{il}^{m}$ and:

\begin{equation}
\{\Pi_{il}^{1},\ldots,\Pi_{il}^{K}\}\propto\{w_{1}p(Y_{il}|\theta_{1}^{*}),\ldots,w_{K}p(Y_{il}|\theta_{K}^{*})\}\label{eq:zetaweights}
\end{equation}
In the above expression, $Y_{il}$ is the data for the $i^{th}$ gene
in class $l$, as mentioned in a previous section. Notice that $z_{il}$
can take any integer value between $1$ and $K$ and that the weights
$\pi_{il}$ depend both on the cluster weights $w_{k}$ and on the
value of the likelihood function $p(Y_{il}|\theta_{k}^{*})$. 

\subsubsection*{Sampling from the conditional posterior $p(W|Z)$}

The mixing weights $W$ are generated using a truncated stick-breaking
process with $\eta=1$. As pointed out in \cite{Ishwaran2001},
this implies that $W$ follows a generalised Dirichlet distribution.
Considering the conjugacy between this and the multinomial distribution,
the first step in updating $W$ is to generate $K-1$ Beta-distributed
random numbers:

\begin{equation}
V_{k}\sim\text{Beta}(1+N_{k},\eta+N-\sum_{m=1}^{k}N_{m})\label{eq:betapost}
\end{equation}
for $k=1,\ldots,K-1$, where $N_{k}$ is the total number of genes
in the $k^{th}$ cluster. Notice that $N_{k}$ can be inferred from
$Z$ by simple counting and $\sum_{m=1}^{K}N_{k}=N$, where $N$ is
the total number of genes. $V_{K}$ is set equal to $1$, in order
to ensure that the weights add up to $1$. These are simply generated
by setting $V_{1}=w_{1}$ and $w_{k}=V_{k}\prod_{m=1}^{k-1}(1-V_{m})$,
as mentioned in a previous section. 

\subsubsection*{Sampling from the conditional posterior $p(\phi|\Theta^{*},Z)$}

The hyper-parameters $\phi=\{a_{\alpha},s_{\alpha},\mu_{\beta},\sigma_{\beta}^{2}\}$
influence indirectly the observations $Y$ through their effect on
the distribution of the active cluster centers, $\Theta_{ac}^{*}=\{\alpha_{ac}^{*},\beta_{ac}^{*}\}$,
where $\alpha_{ac}^{*}=\{\alpha_{ac,1}^{*},\ldots,\alpha_{ac,K_{ac}}^{*}\}$
and $\beta_{ac}^{*}=\{\beta_{ac,1}^{*},\ldots,\beta_{ac,K_{ac}}^{*}\}$.
If we further assume independence between $\alpha_{ac}^{*}$ and $\beta_{ac}^{*}$,
we can write $p(\phi|\Theta^{*},Z)=p(a_{\alpha},s_{\alpha},\mu_{\beta},\sigma_{\beta}^{2}|\alpha_{ac}^{*},\beta_{ac}^{*})=p(a_{\alpha},s_{\alpha}|\alpha_{ac}^{*})p(\mu_{\beta},\sigma_{\beta}^{2}|\beta_{ac}^{*})$. 

Assuming $K_{ac}$ active clusters and considering that the prior
for $\alpha^{*}$ is an Inverse Gamma distribution (see Eq. \ref{eq:thetaprior}),
it follows that the posterior $p(a_{\alpha},s_{\alpha}|\alpha_{ac}^{*})$
is:

\begin{equation}
p(a_{\alpha},s_{\alpha}|\alpha_{ac}^{*})\propto\frac{\gamma_{1}^{a_{\alpha}-1}\exp(-s_{\alpha}\gamma_{2})s_{\alpha}^{a_{\alpha}\gamma_{3}}}{\Gamma(a_{\alpha})^{\gamma_{4}}}\label{eq:aspost}
\end{equation}
The parameters $\gamma_{1}$ to $\gamma_{4}$ are given by the following
expressions:

\begin{eqnarray*}
\gamma_{1} & = & \gamma_{1}^{(0)}\prod_{k=1}^{K_{ac}}\frac{1}{\alpha_{ac,k}^{*}}\\
\gamma_{2} & = & \gamma_{2}^{(0)}+\sum_{k=1}^{K_{ac}}\frac{1}{\alpha_{ac,k}^{*}}\\
\gamma_{3} & = & \gamma_{3}^{(0)}+K_{ac}\\
\gamma_{4} & = & \gamma_{4}^{(0)}+K_{ac}
\end{eqnarray*}
where the initial parameters $\gamma_{1}^{(0)}$, $\gamma_{2}^{(0)}$,
$\gamma_{3}^{(0)}$ and $\gamma_{4}^{(0)}$ are all positive. Since
sampling from Eq. \ref{eq:aspost} cannot be done exactly, we employ
a Metropolis algorithm with acceptance probability

\begin{equation}
P_{acc}=min\left(1,\frac{p(a_{\alpha}^{+},s_{\alpha}^{+}|\alpha_{ac}^{*})}{p(a_{\alpha},s_{\alpha}|\alpha_{ac}^{*})}\right)
\end{equation}
where the proposal distribution $q(\cdot|\cdot)$ for sampling new
candidate points $ $has the same form as in Eq. \ref{eq:propdist}. 

Furthermore, taking advantage of the conjugacy between a Normal likelihood
and a Normal-InverseGamma prior, the posterior probability for parameters
$\mu_{\beta}$ and $\sigma_{\beta}^{2}$ becomes:

\begin{equation}
p(\mu_{\beta},\sigma_{\beta}^{2}|\beta_{ac}^{*})=\text{Normal}\text{InverseGamma}(\delta_{1},\delta_{2},\delta_{3},\delta_{4})\label{eq:musigmapost}
\end{equation}
The parameters $\delta_{1}$ to $\delta_{4}$ (given initial parameters
$\delta_{1}^{(0)}$ to $\delta_{4}^{(0)}$) are as follows:

\begin{eqnarray*}
\delta_{1} & = & \frac{\delta_{1}^{(0)}\delta_{2}^{(0)}+K_{ac}\bar{\beta}_{ac}^{*}}{\delta_{2}^{(0)}+K_{ac}}\\
\delta_{2} & = & \delta_{2}^{(0)}+K_{ac}\\
\delta_{3} & = & \delta_{3}^{(0)}+\frac{K_{ac}}{2}\\
\delta_{4} & = & \delta_{4}^{(0)}+\frac{1}{2}\sum_{k=1}^{K_{ac}}(\beta_{ac,k}^{*}-\bar{\beta}_{ac}^{*})+\frac{1}{2}\frac{\delta_{2}^{(0)}K_{ac}}{\delta_{2}^{(0)}+K_{ac}}(\bar{\beta}_{ac}^{*}-\delta_{1}^{(0)})
\end{eqnarray*}
where $\bar{\beta}_{ac}^{*}=\frac{1}{K_{ac}}\sum_{k=1}^{K_{ac}}\beta_{ac,k}^{*}$.
Sampling a $\{\mu_{\beta},\sigma_{\beta}^{2}\}$-pair from the above
posterior takes place in two simple steps: first, we sample $\sigma_{\beta}^{2}\sim\text{InverseGamma}(\delta_{3},\delta_{4})$,
where $\delta_{3}$ and $\delta_{4}$ are shape and scale parameters,
respectively. Then, we sample $\mu_{\beta}\sim\text{Normal}(\delta_{1},\sigma_{\beta}^{2}/\delta_{2})$. 

\subsection{Algorithm}

We summarise the algorithm for drawing samples from the posterior
$p(\Theta^{*},Z,W,\phi|Y)$ below. Notice that $x^{(t)}$ indicates
the value of $x$ at the $t^{th}$ iteration of the algorithm. $x^{(0)}$
is the initial value of $x$. 

\begin{enumerate}
\item Set $\gamma^{(0)}=\left\{ \gamma_{1}^{(0)},\gamma_{2}^{(0)},\gamma_{3}^{(0)},\gamma_{4}^{(0)}\right\} $
\item Set $\delta^{(0)}=\left\{ \delta_{1}^{(0)},\delta_{2}^{(0)},\delta_{3}^{(0)},\delta_{4}^{(0)}\right\} $ 
\item Set $\phi^{(0)}=\{a_{\alpha}^{(0)}$, $b_{\alpha}^{(0)}$, $\mu_{\beta}^{(0)}$,
$\sigma_{\beta}^{2(0)}\}$ 
\item Set $K$, the truncation level 
\item Sample $\Theta^{*(0)}$ from its prior (Eq. \ref{eq:thetaprior})
conditional on $\phi^{(0)}$
\item Set all $K$ elements of $W^{(0)}$ to the same value, i.e. $1/K$ 
\item Sample $Z^{(0)}$ from the Categorical distribution with weights $W^{(0)}$ 
\item For $t=1,2,3,\ldots,T$

\begin{enumerate}
\item Sample $\Theta_{ac}^{*(t)}$ given $Z^{(t-1)}$, $\phi^{(t-1)}$ and
the data matrix $Y$ using a single step of the Metropolis algorithm
for each active cluster (see Eq. \ref{eq:metropolistheta})
\item Sample $\Theta_{in}^{*(t)}$ from its prior given $\phi^{(t-1)}$
(see Eq. \ref{eq:thetaprior}) 
\item Sample $Z^{(t)}$ given $\Theta^{*(t)}$, $W^{(t-1)}$ and the data
matrix $Y$ (see Eq. \ref{eq:zetaweights})
\item Sample $W^{(t)}$ given $Z^{(t)}$ (see Eq. \ref{eq:betapost}) 
\item Sample $\phi^{(t)}$ given $\Theta_{ac}^{*(t)}$ and $\phi^{(t-1)}$
(see Eqs. \ref{eq:aspost} and \ref{eq:musigmapost}) 
\end{enumerate}
\item Discard the first $T_{0}$ samples, which are produced during the
burn-in period of the algorithm (i.e. before equilibrium is attained),
and work with the remaining $T-T_{0}$ samples.
\end{enumerate}
The above procedure implements a form of blocked Gibbs sampling with
embedded Metropolis steps for impossible to directly sample from distributions. 

\section{Results and Discussion}

We have implemented the methodology described in the preceding sections in software and we have 
applied this software on 
publicly available digital gene expression data (obtained from control and cancerous tissue cultures of neural 
stem cells; \cite{Engstrom:2012ij}) for evaluation purposes. 
The data we used in this study can be found at the 
following URL: http://genomebiology.com/content/supplementary/gb-2010-11-10-r106-s3.tgz.
As shown in Table~\ref{tab:nsdata}, this dataset consists of four libraries from 
glioblastoma-derived neural stem cells and two from non-cancerous neural 
stem cells. Each tissue culture was derived from a different subject. Thus,
the samples are divided in two classes (cancerous and non-cancerous) with four and
two replicates, respectively.

\begin{table}[!t]
\begin{center}
\begin{tabular}{lllllll}\hline
         & \multicolumn{4}{ c }{\textbf{Cancerous}}   & \multicolumn{2}{c}{\textbf{Non-cancerous}} \\
Genes    & GliNS1   & G144      & G166     & G179     & CB541    & CB660    \\ \hline
13CDNA73 & 4        & 0         & 6        & 1        & 0        & 5        \\
15E1.2   & 75       & 74        & 222      & 458      & 215      & 167      \\
182-FIP  & 118      & 127       & 555      & 231      & 334      & 114      \\
\vdots   & \vdots   & \vdots    & \vdots   & \vdots   & \vdots   & \vdots   \\ \hline
\end{tabular}
\caption{Format of the data by \cite{Engstrom:2012ij}. The first four samples are from glioblastoma neural stem cells, while 
the last two are from non-cancerous neural stem cells. The dataset contains a total of 18760 genes (i.e. rows).\label{tab:nsdata}}
\end{center}
\end{table}

We implemented the algorithm presented above in the programming language Python,
using the libraries NumPy, SciPy and MatplotLib. Calculations were
expressed as operations between arrays and the \textit{multiprocessing}
Python module was utilised in order to take full advantage of the
parallel architecture of modern multicore processors. The algorithm was  
run for 200K iterations, which took approximately two days to complete on a 12-core
desktop computer. Simulation results were saved to the disk every 50 iterations. 

\begin{figure}[!tpb]
\centerline{\includegraphics{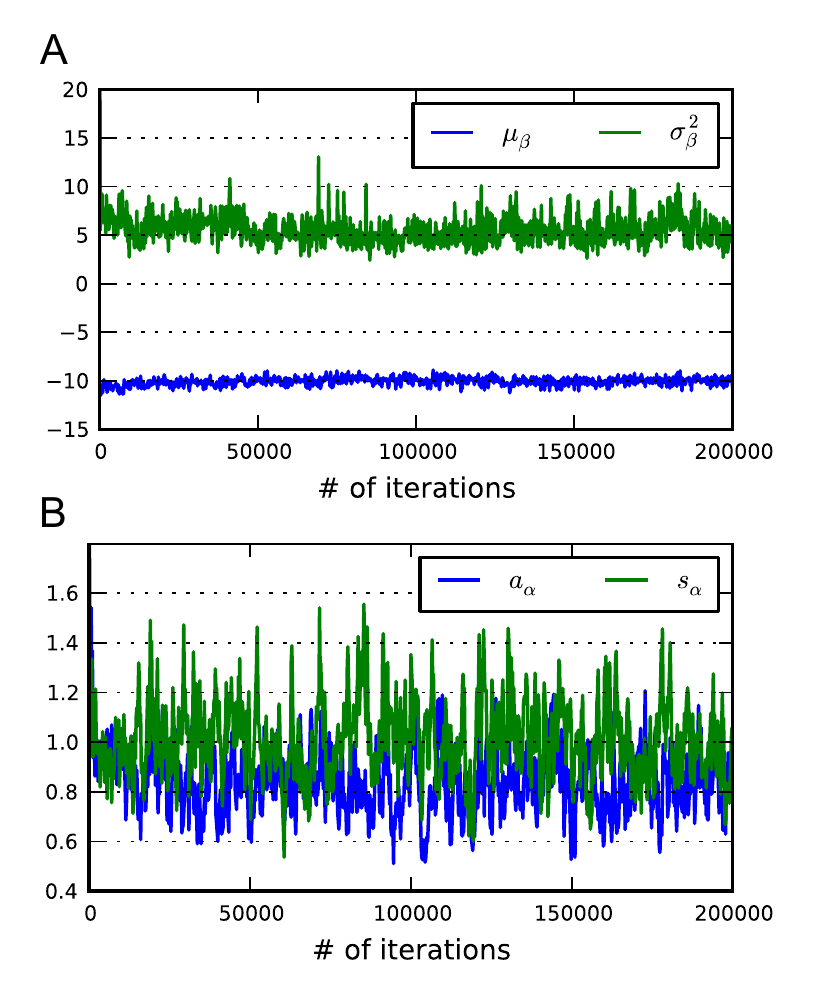}}
\caption{Simulation results after 200K iterations. The chains of random samples correspond
to the components of the vector of hyper-parameters $\phi$, i.e. $\mu_\beta$ and $\sigma^2_\beta$ (panel A)
and $a_\alpha$ and $s_\alpha$ (panel B). The former determines the Normal prior distribution
of the cluster center parameters $\beta^*$, while the latter pair determines the
Inverse Gamma prior distribution of the cluster center parameters $\alpha^*$. The 
random samples in each chain are approximately sampled (and constitute an approximation
of) the corresponding posterior distribution conditional on the data matrix $Y$.}\label{fig:pars}
\end{figure}

The raw simulation output includes chains of random values of 
the hyper-parameters $\phi$, the gene- and class-specific indicators $Z$ and the 
active cluster centers $\Theta_{ac}^*$, which constitute an approximation to the 
corresponding posterior distributions given 
the data matrix $Y$. The chains corresponding to the four different components of 
$\phi=\{a_\alpha,s_\alpha,\mu_\beta,\sigma^2_\beta\}$ are illustrated in Figure~\ref{fig:pars}.    
It may be observed that these reached equilibrium early during the simulation (after less than 20K iterations) 
and they remained stable for the remaining of the simulation. As explained earlier, 
these hyper-parameters are important, because they determine 
the prior distributions of the cluster centers $\alpha^*$ and
$\beta^*$ (hyper-parameters $\{a_\alpha,s_\alpha\}$ and $\{\mu_\beta,\sigma^2_\beta\}$, respectively) and,
subsequently, of the gene- and class-specific parameters $\alpha$ and $\beta$.  

\begin{figure}[!tpb]
\centerline{\includegraphics{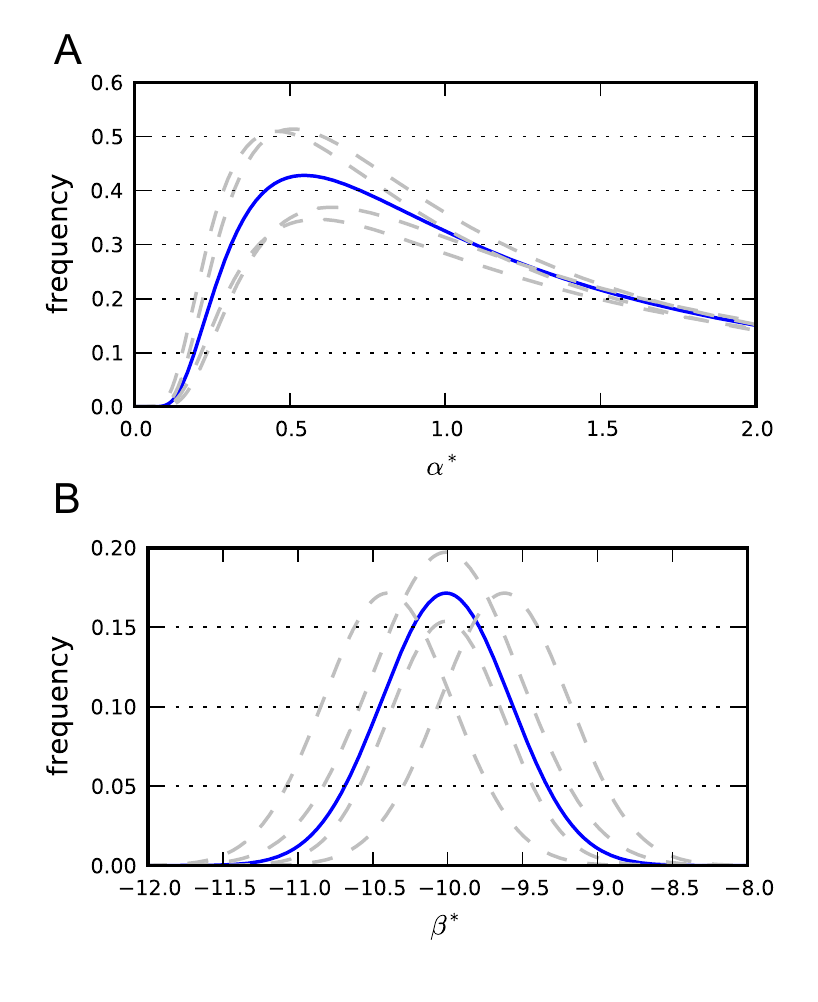}}
\caption{Estimated Inverse Gamma (panel A) and Normal (panel B) prior distributions
for the cluster parameters $\alpha^*$ and $\beta^*$, respectively. The solid lines
indicate mean distributions, i.e. those obtained for the mean values of the hyper-parameters
$a_\alpha$, $s_\alpha$, $\mu_\beta$ and $\sigma^2_\beta$. The dashed lines are distributions  
obtained by adding or subtracting individually one standard deviation from each relevant hyper-parameter.
}\label{fig:dists}
\end{figure}

It follows from analysis of the chains in Figure~\ref{fig:pars} that the estimates 
for these hyper-parameters are (indicating the mean and standard deviation of the estimates): 
$a_\alpha=0.83\pm 0.13$, $s_\alpha=1.00\pm0.16$, $\mu_\beta=-10.01\pm0.39$ and $\sigma^2_\beta=5.41\pm1.32$. 
The corresponding Inverse Gamma and Normal distributions, which are the priors of
the cluster centers $\alpha^*$ and $\beta^*$, respectively, are illustrated in Figure~\ref{fig:dists}.       

\begin{figure}[!tpb]
\centerline{\includegraphics{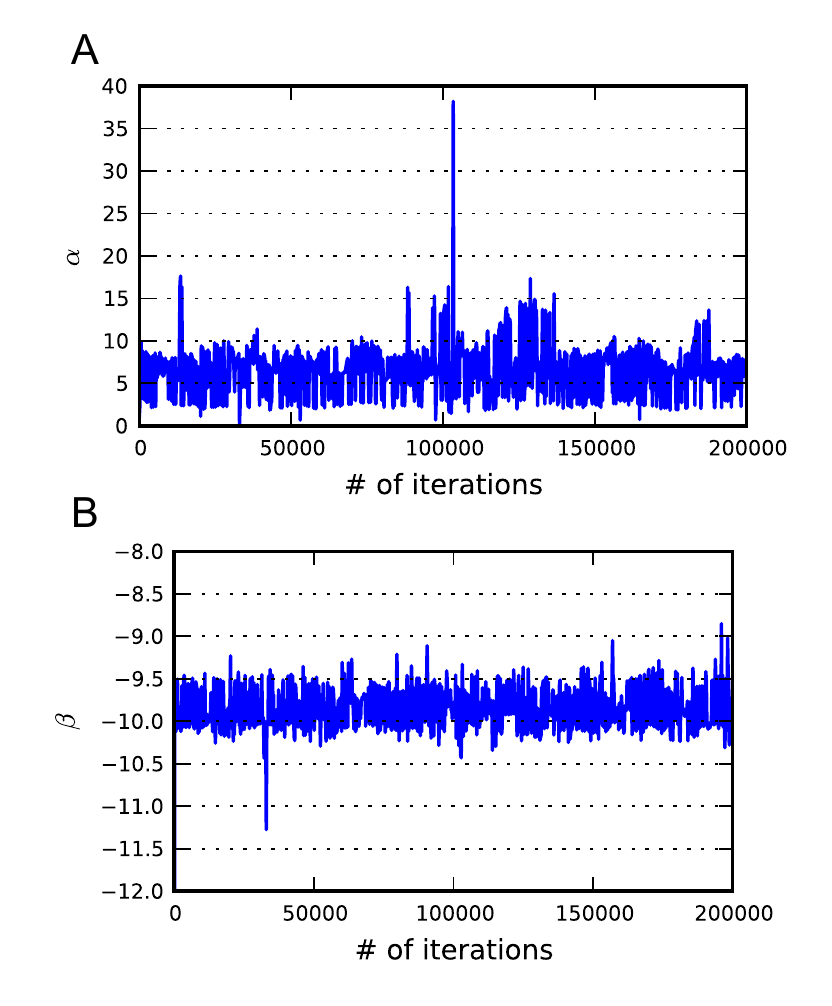}}
\caption{Chains of random samples approximating the posterior distributions of the 
parameters $\alpha$ (panel A) and $\beta$ (panel B) corresponding to the non-cancerous 
class of samples for the tag with ID 182-FIP (third row in Table~\ref{tab:nsdata}). These samples 
were generated after 200K iterations of the algorithm. A similar pair of chains exists
for each gene at each sample class (i.e. cancerous and non-cancerous), although not
all pairs are distinct to each other due to the clustering effect imposed on the data 
by the algorithm.} \label{fig:gene}
\end{figure}

A major use of the methodology presented above is that it allows us to estimate
the gene- and class-specific parameters $\alpha$ and $\beta$, under the assumption  
that the same values for these parameters are shared between different genes or even 
by the same gene among different sample classes. This form of information sharing 
permits pulling together data from different genes and classes for estimating pairs
of $\alpha$ and $\beta$ parameters in a robust way, even when only a small number 
of replicates (or no replicates at all) are available per sample class. 
As an example, in Figure~\ref{fig:gene} we illustrate the chains of random samples 
for $\alpha$ and $\beta$ corresponding to the non-cancerous class of samples for the tag with ID 
182-FIP (third row in Table~\ref{tab:nsdata}). These samples 
constitute approximations of the posterior distributions of the corresponding parameters. 
Despite the very small number of replicates ($n=4$), the variance of the
random samples is finite. Similar chains were derived for each gene in the dataset,
although it should be emphasised that the number of such estimates is smaller than the 
total number of genes, since more than one genes share the same parameter estimates.

\begin{figure}[!tpb]
\centerline{\includegraphics{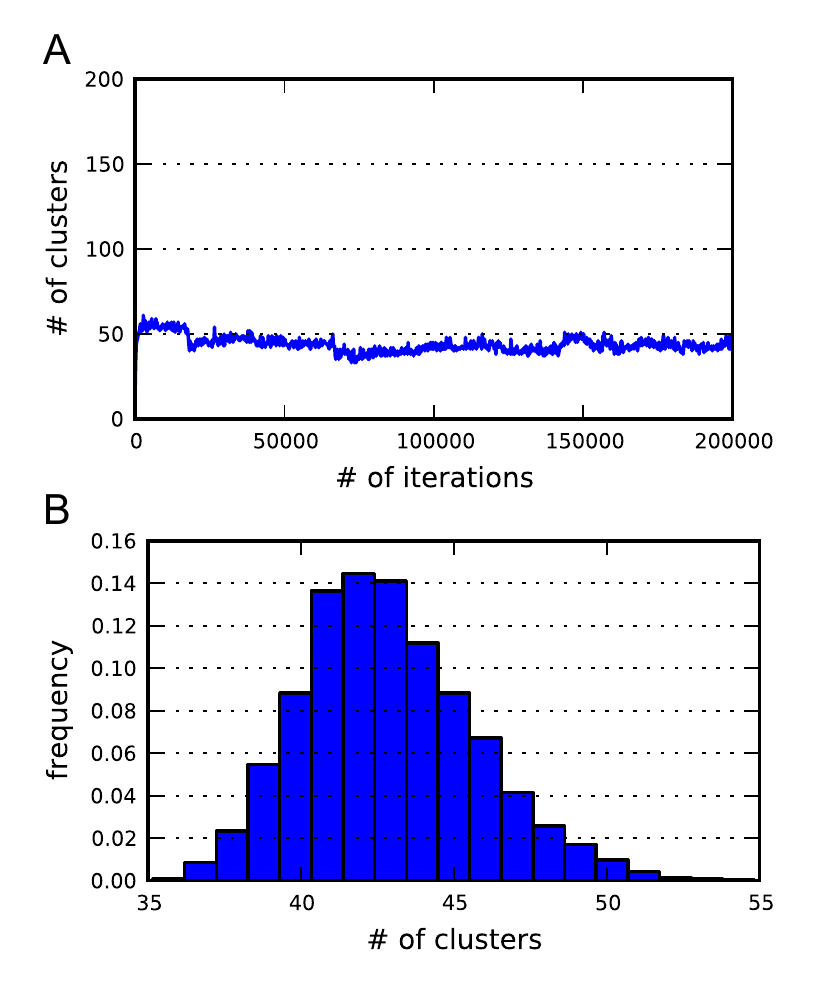}}
\caption{Stochastic evolution of the number of clusters during 200K iterations of the simulation
(panel A) and the resulting histogram after discarding the first 75K iterations as burn-in (panel B).
After reaching equilibrium, the number of clusters fluctuates around a mean of approximately 43
clusters. In general, the estimated number of clusters is much smaller than the truncation
level ($K=200$, see y-axis in panel A). The histogram in panel B approximates the 
posterior distribution of the number of clusters given the data matrix $Y$.}\label{fig:nclusters}
\end{figure}

It has already been mentioned that the sharing of $\alpha$ and $\beta$ parameter 
values between different genes can be viewed as a form of clustering (see Figure~\ref{fig:clustering}), i.e.
there are different groups of genes, where all genes in a particular group
share the same $\alpha$ and $\beta$ parameter values. As expected in a Bayesian
inference framework, the number of clusters is not constant, but it is itself a random
variable, which is characterised by its own posterior distribution and its value 
fluctuates randomly from one iteration to the next. In Figure~\ref{fig:nclusters},
we illustrate the chain of sampled cluster numbers during the course of the simulation
(panel A). The first 75K iterations were discarded as burn-in and the remaining samples
were used for plotting the histogram in panel B, which approximates the posterior 
distribution of the number of clusters given the data matrix $Y$. It may be observed
that the number of clusters fluctuates between 35 and 55 with a peak at around 42
clusters. The algorithm we present above does not make any particular assumptions 
regarding the number of clusters, apart from the obvious one that this number cannot 
exceed the number of genes times the number of sample libraries. Although the 
truncation level $K=200$ sets an artificial limit in the maximum number of clusters, this
is never a problem in practise, since the actual estimated number of clusters is typically
much smaller that the truncation level $K$ (see the y-axis in Figure~\ref{fig:nclusters}A). The 
fact that the number of clusters is not decided a priori, but rather inferred along
with the other free parameters in the model sets the described methodology in an  
advantageous position with respect to alternative clustering algorithms, which require deciding   
the number of clusters at the beginning of the simulation (\cite{Daxin2004}).
 
\begin{figure}[!tpb]
\centerline{\includegraphics{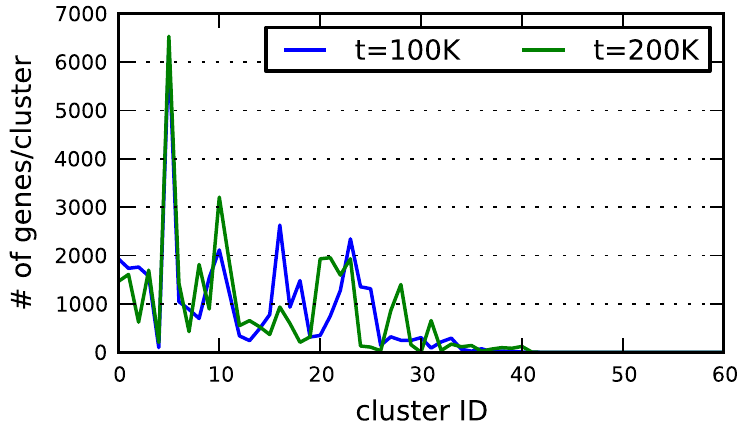}}
\caption{Cluster occupancies after 100K and 200K iterations of the algorithm. 
A single super-cluster (including more then 6000 genes) appears at both stages of the simulation.
The occupancy of the remaining clusters demonstrates some variability during the course 
of the simulation, with clusters containing between 3000 and less than 1000 genes.}\label{fig:occup}
\end{figure}

Similarly to the stochastic fluctuation in the number of clusters, the cluster 
occupancies (i.e. the number of genes per cluster) is a random vector. In Figure~\ref{fig:occup},
we illustrate the cluster occupancies at two different stages of the simulation, i.e.
after 100K and 200K iterations, respectively. We may observe that, with the exception of
a single super-cluster (containing more than 6000 genes), cluster occupancies range from
between around 3000 and less than 1000 genes.   

\begin{figure}[!tpb]
\centerline{\includegraphics{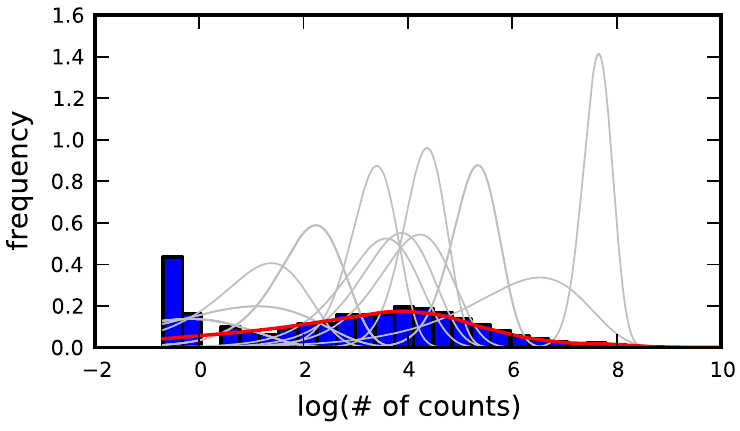}}
\caption{Histogram of the log of the number of reads from sample GliNS1, a subset
of the estimated clusters (gray lines) and the estimated model of the sample at the 
end of the simulation. Each cluster (gray line) represents a Negative Binomial distribution
with specific $\alpha$ and $\beta$ parameters, which models a subset of the count data 
in this particular sample. The complete model
(red line) is the weighted sum of all component clusters.}\label{fig:sample}
\end{figure}

It should be clarified that each cluster includes many (potentially, hundreds of) genes 
and it may span several classes. An individual cluster represents a Negative Binomial
distribution (with concrete $\alpha$ and $\beta$ parameters), which models with high
probability the count data from all its member genes. This is illustrated in 
Figure~\ref{fig:sample}, where we show the histogram of the log of the count data
from the first sample (sample GliNS1 in Table~\ref{tab:nsdata}) along with a subset 
of the estimated clusters after 200K iterations (gray lines) and the fitted model
(red line). It may be observed that each cluster models a subset of the gene expression
data in the particular sample. The complete model describing the whole sample is
a weighted sum of the individual clusters/Negative Binomial distributions. Formally,
\begin{eqnarray}
p(Y_j|\alpha_{1\lambda(j)},\beta_{1\lambda(j)},\ldots,\alpha_{N\lambda(j)},\beta_{N\lambda(j)})= 
\frac{1}{N}\sum_{i=1}^N p(y_{ij}|\alpha_{i\lambda(j)},\beta_{i\lambda(j)})
\end{eqnarray}
\noindent where $Y_j$ is the $j^{th}$ sample and the index $i$ runs over all $N$ genes.
We repeat that not all $\{\alpha_{i\lambda(j)},\beta_{i\lambda(j)}\}$ pairs are distinct.
Also, clusters with larger membership (i.e. including a larger number of genes) have larger
weight in determining the overall model.

\begin{figure}[!tpb]
\centerline{\includegraphics{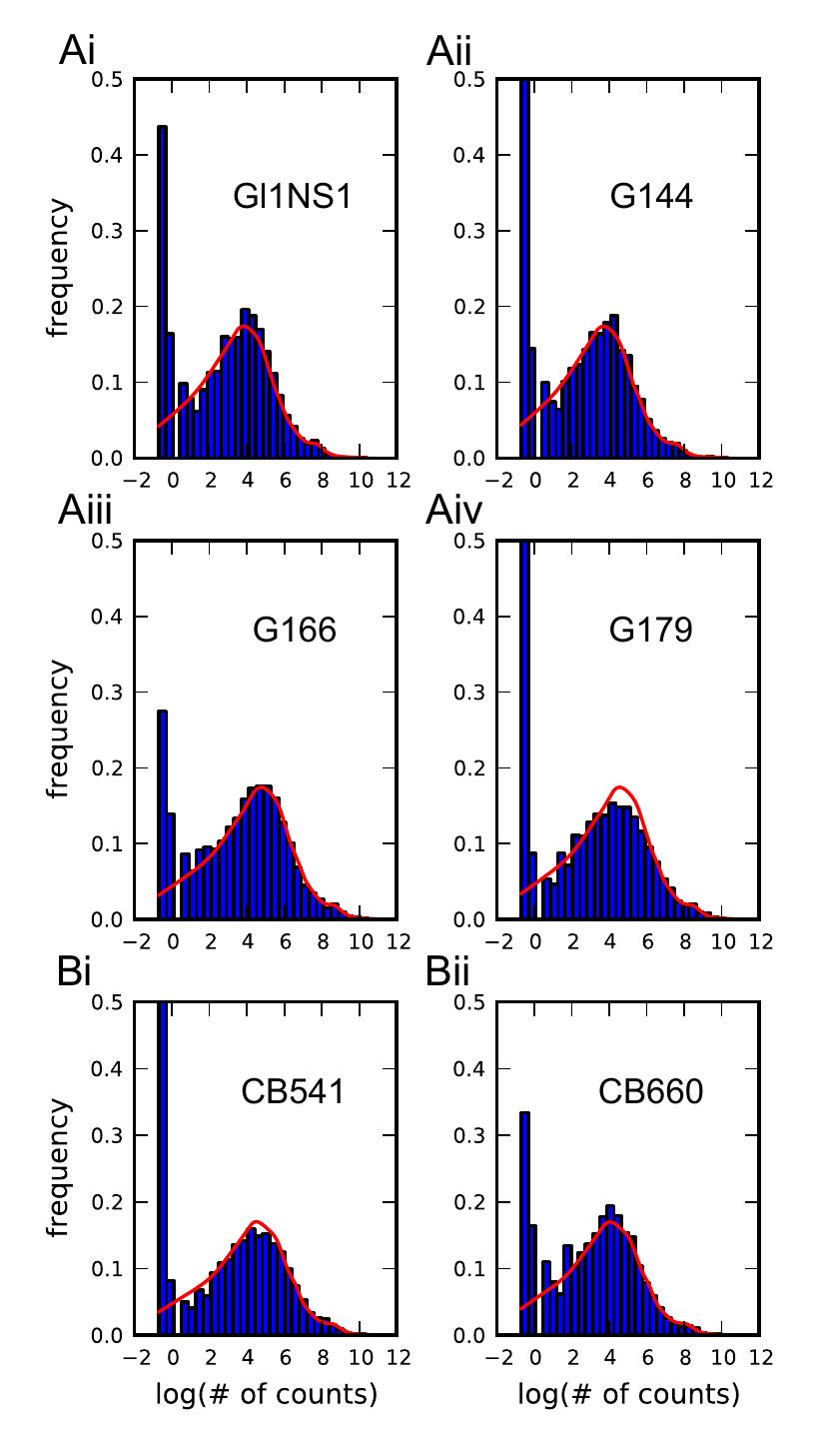}}
\caption{Histograms of the log of the number of reads from cancerous (panels Ai-iv) and
non-cancerous (panels Bi,ii) samples and the respective estimated models after 200K 
iterations of the algorithm. As already mentioned, each red line is the weighted sum
of many component Negative Binomial distributions / clusters, which model different 
subsets of each data sample. We may observe that the estimated models fit tightly 
the corresponding data samples. }\label{fig:samples}
\end{figure}

The proposed methodology provides a compact way to model 
each sample in a digital gene expression dataset following a two-step procedure:
first, the dataset is partitioned into a finite number of clusters, where each cluster
represents a Negative Binomial distribution (modelling a subset of the data) and the parameters 
of each such distribution are estimated. Subsequently, each
sample in the dataset can be modelled as a weighted sum of Negative Binomial distributions.
In Figure~\ref{fig:samples}, we show the log of count data for each sample 
in the dataset shown in Table~\ref{tab:nsdata} along with the fitted models (red lines)
after 200K iterations of the algorithm.

%
%

\section{Conclusion}

Next-generation sequencing technologies are routinely being used for generating huge volumes 
of gene expression
data in a relatively short time. This data is fundamentally discrete in nature and
their analysis requires the development of novel statistical methods, rather than
modifying existing tests that were originally aimed at the analysis of microarrays.       
The development of such methods is an active area of research and several papers have been 
published on the subject (see \cite{Oshlack:2010vn} and \cite{Kvam:2012hc} for an overview).

In this paper, we present a novel approach for modelling over-dispersed
count data of gene expression (i.e. data with variance larger than the mean predicted
by the Poisson distribution) using a hierarchical model based on the Negative Binomial
distribution. The novel aspect of our approach is the use of a Dirichlet process 
in the form of stick breaking priors for modelling the parameters (mean and over-dispersion) 
of the Negative Binomial distribution. By construction, this formulation forces
clustering of the count data, where genes in the same cluster are sampled
from the same Negative Binomial distribution, with a common pair of mean and over-dispersion
parameters. Through this elegant form of information sharing between genes, we compensate
for the problem of little or no replication, which often restricts the analysis of digital gene expression
datasets. We have demonstrated the ability of this approach to model accurately
actual biological data by applying the proposed methodology on a publicly available dataset
obtained from cancerous and non-cancerous cultured neural stem cells (\cite{Engstrom:2012ij}).

We show that inference is achieved in the proposed model through the application of a blocked
Gibbs sampler, which includes estimating, among others, the
gene- and class-specific mean and over-dispersion of the Negative Binomial distribution.
Similarly, the number of clusters and their occupancies are inferred along with
the rest free parameters in the model.

Currently, the software implementing the proposed method remains relatively computationally expensive.
In particular, 200K iterations require approximately two days to complete on a 12-core desktop computer. 
This time scale is not disproportionate to the production time of experimental data and it 
is mainly due to the high volume of the tested data ($>15K$ genes per sample) and the need to 
obtain long chains of samples for a more accurate estimation of posterior distributions.   
Long execution times are a characteristic, more generally, of all Monte Carlo approximation methods.
Our implementation of the algorithm is completely parallelised
and calculations are expressed as operations between vectors in order to take full
advantage of modern multi-core computers. Ongoing work towards reducing execution 
times aims at the application of variational inference methods (\cite{Blei2006}), 
instead of the blocked Gibbs sampler we currently use. The algorithm can be further improved by avoiding
truncation of the infinite summation described in Equation~\ref{eq:infmix}, as 
described in \cite{Papaspiliopoulos2008} and in \cite{Walker2007}.             
        
This non-parametric Bayesian approach for modelling count data has thus shown great promise 
in handling over-dispersion and the all-too-common problem of low replication,
both in theoretical evaluation and on the example dataset. 
The software that has been produced will be of great utility for the study of digital gene 
expression data and the statistical theory will contribute to leading the development 
of non-parametric methods in general for all forms of modelling count data of gene expression.

\section*{Acknowledgement}
The authors would like to thank Prof. Peter Green and Dr. Richard Goldstein for useful discussions.
Also, we would like to thank P. G. Engstrom and colleagues for producing the public
data we used in this paper.  

\paragraph{Funding:} This work was supported by grants EPSRC EP/H032436/1 and BBSRC G022771/1.

\bibliographystyle{plain}
\bibliography{document}

\end{document}